\documentclass[12pt,twoside]{article}

\usepackage{amsmath,amsfonts}
\usepackage{algorithmic}
\usepackage{algorithm}
\usepackage{array}
\usepackage[caption=false,font=normalsize,labelfont=sf,textfont=sf]{subfig}
\usepackage{booktabs}
\usepackage{textcomp}
\usepackage{stfloats}
\usepackage{url}
\usepackage{verbatim}
\usepackage{graphicx}
\usepackage{cite}
\usepackage{bm} 
\usepackage{multirow}
\usepackage{amssymb}
\usepackage{pifont}
\usepackage[table,xcdraw]{xcolor}
\usepackage{booktabs}
\definecolor{Gray}{gray}{0.9}
\usepackage[colorlinks=true, linkcolor=blue, citecolor=green, urlcolor=cyan]{hyperref}

\def\*#1{\mathbf{#1}}
\def\+#1{\mathcal{#1}}
\def\-#1{\mathbb{#1}}
\def\~#1{\mathrm{#1}}
\def\R{\mathbb{R}}

\usepackage{pgfplots}
\usepackage{pgf}
\usepackage{tikz}
\usetikzlibrary{matrix, calc}
\usetikzlibrary{positioning, arrows.meta, spy}

\usetikzlibrary{fit}					  
\usetikzlibrary{backgrounds}	         
\usetikzlibrary{shapes.arrows}
\usetikzlibrary{shapes.geometric}
\usetikzlibrary{shapes.multipart}
\usetikzlibrary{matrix, arrows, positioning}
\tikzstyle{rec}=[draw,rectangle, minimum height=2cm]
\tikzset{>=stealth', punkt/.style={rectangle, 
		fill=gray!40, 
		draw=black, very thick, text width=5.2em, minimum height=2.5em, text centered}}
\tikzset{>=stealth', Denoi/.style={rectangle, fill=blue!20, 
		draw=black, very thick, text width=6em, minimum height=3.5em, text centered}}
\tikzset{>=stealth', CG/.style={rectangle, 
		fill=green!20, draw=black, very thick, text width=9.5em, minimum height=3.5em, text centered}}
\tikzstyle{background} = [rectangle, fill=green!20, inner sep=0.1cm, rounded corners=4mm, 
minimum height=6em]
\tikzstyle{sum}   = [draw, fill=gray!40, circle, node distance=1cm]
\tikzstyle{dot}   = [circle, fill=black, inner sep=0pt, minimum size=5pt, node contents={}]
\tikzstyle{fig_n} = [node distance=30pt, inner sep=0cm]

\usepackage[super,sort,comma]{natbib}

\usepackage{fancyhdr}

\usepackage[section]{placeins}   

\usepackage{graphicx}
\usepackage{makecell}

\makeatletter \renewcommand\@biblabel[1]{$^{#1}$} \makeatother
\newcommand{\cen}[1]{\begin{center} #1 \end{center}}

\usepackage{hyperref}
\hypersetup{ colorlinks,
    citecolor=blue,
    filecolor=blue,
    linkcolor=blue,
    urlcolor=blue
}

\usepackage{xcolor}
\definecolor{gray}{rgb}{0.6,0.6,0.6}
\definecolor{red}{rgb}{0.85,0,0}
\definecolor{green}{rgb}{0,0.85,0}
\definecolor{blue}{rgb}{0,0,0.85}
\definecolor{beige}{rgb}{0.92,0.87,0.78}
\usepackage[all]{hypcap}   

\begin{document}

\cen{\sf {\Large {\bfseries Equivariant Conditional Diffusion Model for Head and Neck CT Image Synthesis from CBCT}  \\  
\vspace*{8mm}
Alzahra Altalib$^{1,2}$, Chunhui Li$^1$, Alessandro Perelli$^3$} \vspace*{5mm}\vspace{2mm}\\
$^1$ School of Science and Engineering, University of Dundee, DD1 4HN, UK.\vspace{2mm}\\
$^2$ Department of Applied Medical Sciences, Jordan University of Science and Technology, Irbid, 21410, Jordan \vspace{2mm}\\
$^3$ School of Cardiovascular and Metabolic Health,
College of Medicine, Veterinary and Life Sciences, University of Glasgow, G12 8TA, UK\\
}

\pagenumbering{roman}
\setcounter{page}{1}
\pagestyle{plain}

Corresponding author: Alzahra Altalib. email: \url{2600129@dundee.ac.uk} \\

\begin{abstract}
	\noindent {\bf Background:} Cone-beam computed tomography (CBCT) is a commonly used modality for image guided radiotherapy (IGRT). It offers real time anatomical visualization with low acquisition cost and dose. Nevertheless, photon scattering and beam hindrance lead CBCT images to suffer from several artifacts. These involve inaccurate Hounsfield Unit (HU) values, which render a lower reliability towards the dose calculations and adaptive planning.  \\
    {\bf Purpose:} Computed tomography (CT), on the contrary, offers better image quality and accurate HU calibration, yet is typically acquired using offline mode and fails to capture the intra-treatment anatomical changes. This renders a need for developing an accurate CBCT to CT synthesis to mitigate the gap in imaging quality in the adaptive radiotherapy workflow. \\
    {\bf Methods:} To cater to this, we propose a novel diffusion based conditional generative model, coined EqDiff-CT, to synthesize high quality CT images from CBCT. EqDiff-CT employs a denoising diffusion probabilistic model (DDPM) to iteratively inject noise and learn latent representations that enable reconstruction of anatomically consistent CT images. A group equivariant conditional U-Net backbone, implemented with e2cnn steerable layers, enforces rotational equivariance (cyclic C4 symmetry), helping preserve fine structural details while minimizing noise and artifacts. \\
    {\bf Results:} The system was trained and validated on the SynthRAD2025 dataset, comprising CBCT–CT scans across multiple head and neck anatomical sites, and we compared it with advanced methods such as CycleGAN and DDPM. EqDiff-CT provided substantial gains in structural fidelity, HU accuracy and quantitative metrics. Visual findings further confirm the improved recovery, sharper soft tissue boundaries, and realistic bone reconstructions. \\
    {\bf Conclusions:} The findings suggest that the diffusion model has offered a robust and generalizable framework for CBCT improvements. The proposed solution helps in improving the image quality as well as the clinical confidence in the CBCT guided treatment planning and dose calculations. 

\end{abstract}

\newpage

\setlength{\baselineskip}{0.7cm}      	

\pagenumbering{arabic}
\setcounter{page}{1}
\pagestyle{fancy}

\section{Introduction}

Cone-Beam Computed Tomography (CBCT) and conventional fan-beam Computed Tomography (CT) serve as integral modalities in image guided radiotherapy (IGRT) \cite{int1,int2}. CBCT enables volumetric acquisition through a single rotation and is therefore frequently used \cite{cbct1,cbct2,cbct3}; it offers high spatial resolution and integrates conveniently with linear accelerators for daily image guidance. This supports accurate patient setup and adaptation to interfractional anatomical changes, facilitating adaptive radiotherapy (ART). CT, on the other hand, provides better soft tissue contrast and a high signal to noise ratio (SNR) with accurate Hounsfield Unit (HU) calibration \cite{ct1,ct2}, enabling reliable dose calculation and tissue characterization. Despite these strengths, CBCT suffers from increased scatter, truncated projections, and inconsistent HU values due to artifacts and non standard calibration \cite{cbct4,cbct5}. CT is typically acquired offline during planning and thus cannot reflect day to day anatomical variations during treatment \cite{ct1,ct2}. 

To address these limitations, synthetic CT (sCT) generation from CBCT has emerged as a promising strategy. To address these limitations, synthetic CT (sCT) generation from CBCT is widely investigated. The goal is to map CBCT images to CT like image quality, yielding HU consistent, artifact reduced volumes \cite{sCT1,sCT2}. This can improve dose calculations and anatomical monitoring and can facilitate online or offline ART, by enabling plans based on daily anatomy without the logistical burden of acquiring repeat CT scans. 

From a clinical perspective, HU accuracy in sCT is essential, as uncertainties in electron density mapping can lead to dose discrepancies of several percent, which may compromise target coverage or increase normal tissue toxicity \cite{lavrova2023}. Several studies have demonstrated that sCT based recalculations achieve dose distributions within 1–2\% of reference CT, underscoring their reliability for adaptive workflows. In parallel, high quality sCT volumes enable more consistent segmentation of targets and organs at risk compared to raw CBCT, reducing inter observer variability and improving the robustness of auto segmentation algorithms \cite{miller2019}. These advances have direct clinical impact by supporting accurate treatment adaptation, minimizing geographic misses, and ensuring safe dose escalation when indicated. 

sCT generation is particularly relevant for head and neck, pelvic, and thoracic sites, where anatomical changes are frequent and dosimetric accuracy is critical. In addition, sCT generation reduces imaging dose and can streamline workflows while improving patient comfort \cite{sCT3,sCT4}.

Several methods have been explored in the context of CBCT to CT synthesis \cite{altalib2025}. The traditional methods for CBCT enhancement and sCT generation rely on deformable image registration (DIR) and analytical intensity correction. These methods are primarily intended to deform planning CT (pCT), or the reference CT images into the geometry of daily CBCT scans to enable dose recalculation \cite{1,7,24,25}. While fast and training free, they depend on prior CT anatomy and are unable to recover high frequency structures, limiting adaptability to large anatomical variations. These limitations motivate data driven methods that learn CT appearance directly from CBCT.

GANs have become pivotal in CBCT to CT synthesis, especially in the CycleGAN variants which are widely used in CBCT→CT synthesis with unpaired training. This is due to their ability to learn mappings from unpaired image domains. The application of such models involve pelvic \cite{4,16}, thoracic \cite{6,peng2024conditional}, abdominal \cite{8,23}, and H\&N imaging \cite{14,19}. Domain adapted and attention augmented variants have demonstrated improved robustness in the presence of anatomical variability \cite{14,22}. Pediatric studies have also reported acceptable clinical accuracy \cite{23}, while other studies \cite{24,25} have compared CycleGAN results with commercial DIR/AIC pipelines, revealing improved HU accuracy and better dose conformity. However, adversarial training can be unstable, with risks of mode collapse and hallucinated structures, and interpretability remains a concern for clinical deployment.

Some of the multi model comparisons that have been explored involve cGANs, UNets, and hybrid approaches. It has been established that cGANs have outperformed other models in MAE and Dice coefficient for nasopharyngeal imaging \cite{16}, however, GANs suffer from instability in training. In addition, the hallucination artifacts and lack of interpretability limit their applications in clinical settings. Therefore, a need for the development of a model that can offer spatial consistency exists in long range context modelling.

To overcome the limitations associated with GANs, several CNN based models have been explored. These generally rely on U Net backbones with residual connections, attention blocks, or transformers. For instance, a multiresolution residual network has been proposed that reduces MAE and improves SSIM in pelvic CBCT \cite{15}. Similarly, a ResNet with perceptual loss achieved high PSNR in pelvic imaging \cite{17}. Transformer based methods, including Swin Transformer U-Net, capture long range spatial features in abdominal datasets \cite{18}. ResUNet with self attention has outperformed traditional CNNs in H\&N sCT synthesis while preserving critical anatomy \cite{33}. A dual cycle GAN with patch attention has also been proposed for thoracic sCT synthesis, improving MAE and spatial consistency \cite{31}.

Some hybrid architectures have been developed in this context. For instance, VoxelMorph GAN combines deformable registration with generative learning for improved alignment and anatomical accuracy using abdominal data \cite{32}. DenseUNet and attention CNN models have utilized joint losses including MAE, adversarial, and perceptual terms to attain low contrast abdominal and thoracic sCT outcomes \cite{20,thummerer2020}. This architecture aids robustness and generalizability on unseen data, yet interpretability and real time execution remain challenging. These limitations suggest a need for more stable and probabilistically sound generative models.

In recent times, diffusion models have been explored in medical image synthesis \cite{altalib2025cond}. These models offer improved training stability, sample diversity, and strong theoretical grounding. Li et al.\ \cite{li2023frequency} proposed a frequency guided diffusion model (FGDM) with high/low pass frequency regularizations that enhanced anatomical fidelity during domain translation. Sun et al.\ \cite{sun2024stacked} proposed a coarse to fine hierarchical diffusion model that refined image quality via stacked denoising stages. Patient specific fine tuning has been investigated as a viable strategy in \cite{chen2024patient} and \cite{peng2024conditional}, tailoring Denoising Diffusion Probabilistic Model (DDPM) to individualized anatomical distributions for improved structural consistency in lung and head and neck regions. Different input representations for CBCT to CT synthesis Conditional DDPM have been explored in \cite{altalib2026}. Further studies explore Swin U-NET backbones \cite{viar2024dual}, hybrid frequency embeddings \cite{yin2024latent}, and dual branch attention networks \cite{zhang2024texture}, contributing to texture preservation and dosimetric accuracy. Although these studies serve as a good baseline for diffusion models for sCT generation, limitations persist. Computational cost at training/inference remains a bottleneck, and generalization across anatomical sites and frequency domain variability is under explored. 

\subsection{Contribution of This Work}
In comparison to the existing registration based and adversarial methods, this work presents a conditional denoising diffusion framework. The method has been designed for sCT images generation, where the mapping between CBCT and CT is learned without the need for deformable priors or adversarial training. The method makes use of a time conditioned, group equivariant U-Net denoiser (via \texttt{e2cnn}) that operates on the 2D axial slices with discrete rotational equivariance (cyclic C4 symmetry). The self attention blocks have been integrated for capturing the long range spatial dependencies. The training part minimizes the non adversarial hybrid objective that combines mean square error (MSE) and a structural similarity index (SSIM). These are adopted on the predicted noise, and the inference employs the variance correct reverse diffusion. The evaluations have been conducted on the SynthRAD2025 Head \& Neck dataset \cite{Thummerer2025SynthRAD2025}. The patient wise split of 80/20 has been used while making both slices and volume wise metrics to be reported. The contributions of the work thus include 1) a conditional DDPM framework for CBCT to CT synthesis that performs slice wise synthesis with volume wise analysis; 2) a group equivariant, attention enhanced, time conditioned U-Net denoiser (via \texttt{e2cnn}) with CBCT concatenation at every time step, ensuring orientation consistent reconstructions and 3) a stable, non adversarial training objective combining pixel wise losses (MSE and SSIM) on predicted noise. This is carried out along with an HU preserving preprocessing/post processing pipeline. 

Several experiments have been conducted on the SynthRAD2025 (Head \& Neck) that demonstrated that the attention enhanced diffusion model outperforms both a baseline diffusion model (without attention) and a CycleGAN baseline across SSIM, PSNR, MSE, and MAE. In addition, the model preserved the clinically relevant structures, including the mandible, airway, and cervical spine.

\subsection*{Notation and Organization of the Paper}
We adopt the following notations throughout the
manuscript: discrete operators or matrices and column vectors
are written, respectively, in capital and normal boldface type, i.e., $\*A$ and $\*a$, to distinguish from scalars and continuous variables written in normal weight; an image $\*x\in\R^{N\times N}$ is represented by a matrix for algebraic operations. The specific variables used in the definition of the EqDiff-CT algorithm are the following: 

\begin{itemize}
	\item $G$: rotation group $C_4$.
	\item $R_g$: spatial action of $g\in G$ on $\mathbb{Z}^2$.
	\item $\bm\rho_{\mathrm{in}}, \bm\rho_{\mathrm{out}}$: input/output channel representations.
	\item $f, z^{(l)}$: feature fields (tensors over spatial grid with channel type).
	\item $\kappa$: steerable kernel obeying $\rho$-constraints.
	\item $\Phi$: equivariant linear operator. 
	\item \texttt{InnerBatchNorm}: equivariant normalization.
	\item \texttt{Norm-ReLU}: norm-based gated nonlinearity.
	\item $\*Q,\*K,\*V$: equivariant projections for attention.
	\item $\*x_t, \*x^c, t, \bm\epsilon_{\bm\theta}$: DDPM variables (noisy input, condition, time step, score network).
	\item $\beta_t,\alpha_t,\bar\alpha_t,\sigma_t$: diffusion schedule parameters.
\end{itemize}

Finally, the expectation respect
to random variables $a, b$ is indicated with the notation $\-E_{a,b}$. The structure of this article is organized as follows: in Section~\ref{sec:method} we introduce the Conditional Diffusion model framework. Section~\ref{sec:equiv} describes our proposed EqDiff-CT method for CT image synthesis and Section~\ref{sec:dataset} details the implementation aspects for the clinical dataset preparation. Section~\ref{sec:results} introduces the common settings for training with the real clinical dataset SynthRAD2025, the ablation study and it shows the experimental results compared with other state of the art deep learning methods for image synthesis. Section~\ref{sec:conclusion} provides a summary of the results with the proposed EqDiff-CT method and future work.

\section{Methods}\label{sec:method}
In this section, the methodology of the Conditional Diffusion Probabilistic model adopted for the development of the EqDiff-CT framework is presented. 

\subsection{Conditional Diffusion Model for CBCT to CT Synthesis}

The generative mechanism has been developed using DDPM formalization. The reference CT image $\*x_0\in\R^{H\times W}$ is passed on through the model and progressively corrupted using a forward stochastic process. The generative model $\epsilon_{\bm\theta}$ subsequently learns to denoise and reconstruct $\*x_0$ from noise. Overall, the model is conditioned on the paired CBCT image $\*x^c\in\R^{H\times W}$, which enables a conditional generation via $\hat{\*x}_0 \sim p_{\bm\theta}(\*x_0 | \*x^c)$. 
The forward and reverse diffusion process used in this study follows the standard DDPM formulation \cite{ho2020ddpm}, while the encoder  decoder
backbone is based on the U-Net architecture~\cite{ronneberger2015} with
self attention components inspired by prior attention based neural network models~\cite{vaswani2017attention}. 

\subsubsection{Forward Diffusion Process}
Let $\{\beta_t\}_{t=1}^T$ be a monotonically increasing variance schedule which is linearly spaced over $[{\beta}_1, \beta_T]$. The forward process is a fixed Markov chain:
\begin{equation}
	q(\*x_t | \*x_{t-1}) = \mathcal{N}(\*x_t; \sqrt{1 - \beta_t} \*x_{t-1}, \beta_t \mathbf{I})
\end{equation}
By recursive application, the process has been marginalized at arbitrary timestep $t$ as:
\begin{equation}
	q(\*x_t | \*x_0) = \mathcal{N}(\*x_t; \sqrt{\bar{\alpha}_t} \*x_0, (1 - \bar{\alpha}_t) \mathbf{I})
\end{equation}
where $\bar{\alpha}_t = \prod_{s=1}^t (1 - \beta_s)$. 
The noise addition has been simulated during the training phase as follows:
\begin{equation}
	\*x_t = \sqrt{\bar{\alpha}_t} \*x_0 + \sqrt{1 - \bar{\alpha}_t} \bm\epsilon, \quad \bm\epsilon \sim \mathcal{N}(0, \mathbf{I})
\end{equation}

\subsubsection{Time Conditional Denoising Objective}

The model $\epsilon_{\bm\theta}\left(\*x_t, \*x^c, t\right)$ has been trained to predict the noise component $\epsilon$. This is added to the ground truth CT image $\*x_0$ and conditioned on a CBCT image $\*x^c$ that is concatenated channel wise. The loss function is therefore formulated as a denoising score matching objective:
\begin{equation}
	\mathcal{L}_{\text{DDPM}}(\bm\theta) = \mathbb{E}_{\*x_0, \epsilon, t} \left[ 
	\left\| \epsilon - \epsilon_{\bm\theta}(\*x_t \| \*x^c, t) \right\|_2^2 
	\right]
\end{equation}
where $\*x_t \| \*x^c$ denotes the concatenation of the noisy CT with the unperturbed CBCT slice across the channel dimension.

\subsubsection{Reverse Process and Sampling}

The generative sampling commences at standard Gaussian noise $\*x_T \sim \mathcal{N}(0, \mathbf{I})$ and is recursively progressed to compute posterior approximations leading to the reverse diffusion trajectory:
\begin{equation}
	p_{\bm\theta}(\*x_{t-1} | \*x_t, \*x^c) = \mathcal{N}\left(\mu_\theta(\*x_t, \*x^c, t), \sigma_t^2 \mathbf{I}\right)
\end{equation}
The mean is further computed from the predicted noise by using \cite{song2021}:
\begin{equation}
	\mu_{\bm\theta}(\*x_t, \*x^c, t) = \frac{1}{\sqrt{\alpha_t}} \left( \*x_t - \frac{1 - \alpha_t}{\sqrt{1 - \bar{\alpha}_t}} \cdot \epsilon_{\bm\theta}\left(\*x_t \| \*x^c, t\right) \right)
\end{equation}
The variance $\sigma_t^2$ is either fixed or learned and subsequently uses the closed form posterior variance derived from the forward chain:
\begin{equation}
	\sigma_t^2 = \beta_t \cdot \frac{1 - \bar{\alpha}_{t-1}}{1 - \bar{\alpha}_t}
\end{equation}
This process has been iterated from $t = T$ to $t = 0$ for the generation of a high fidelity sCT image, which is conditioned on the CBCT input. The final output is rescaled back to the clinical HU range by using inverse normalization.

\subsubsection{Conditional Sampling Stability}
To ensure robust sampling, Gaussian noise has been injected and scaled by $\sqrt{\sigma_t^2}$ at each step $t > 0$, and omitted at $t = 0$:
\begin{equation}
	\*x_{t-1} = \mu_{\bm\theta}(\*x_t, \*x^c) + \sqrt{\sigma_t^2} \,\bm\epsilon, \quad \bm\epsilon \sim \mathcal{N}(0, \mathbf{I})
\end{equation}
This ensures that variance-correct sampling takes place along with preserving the conditioning information from CBCT at each timestep.

\subsubsection{Training Objective}

The proposed conditional diffusion framework has been trained on paired CBCT CT images. It incorporates a combination of a stochastic forward process and sampling for conditional denoising predictions. The ultimate objective is associated with the perceptually aligned loss. 

Let $\*x_0$ represent the ground truth CT slice and $\*x^c$ the corresponding CBCT slice. At each iteration, a timestep $t \sim \mathcal{U}(0, T-1)$ has been randomly sampled. A noisy version $\*x_t$ is generated using the forward process $q(\*x_t | \*x_0)$ as:
\begin{equation}
	\*x_t = \sqrt{\bar{\alpha}_t} \*x_0 + \sqrt{1 - \bar{\alpha}_t} \bm\epsilon, \quad \bm\epsilon \sim \mathcal{N}(0, \*I)
\end{equation}
The denoising model $\epsilon_{\bm\theta}(\*x_t, \*x^c, t)$ then determined the original noise $\bm\epsilon$ that was used to perturb $\*x_0$.

In addition to the standard Mean Squared Error (MSE) objective:
\begin{equation}
	\mathcal{L}_{\text{MSE}} = \mathbb{E}_{\*x_0, \bm\epsilon, t} \left[
	\left\| \epsilon - \epsilon_{\bm\theta}(\*x_t, \*x^c, t) \right\|_2^2
	\right]
\end{equation}
a perceptual loss has been introduced by using SSIM as follows:
\begin{equation}
	\mathcal{L}_{\text{SSIM}} = 1 - \text{SSIM}\left(\epsilon, \epsilon_{\bm\theta}(\*x_t, \*x^c, t)\right)
\end{equation}
The final hybrid loss is computed as a weighted sum:
\begin{equation}
	\mathcal{L}_{\text{hybrid}} = \lambda_{\text{MSE}} \cdot \mathcal{L}_{\text{MSE}} + \lambda_{\text{SSIM}} \cdot \mathcal{L}_{\text{SSIM}}
\end{equation}
where $\lambda_{\text{MSE}}$ and $\lambda_{\text{SSIM}}$ are scalar values chosen a priori. 

\begin{figure}[!h]
	\centering
\includegraphics[width=\linewidth]{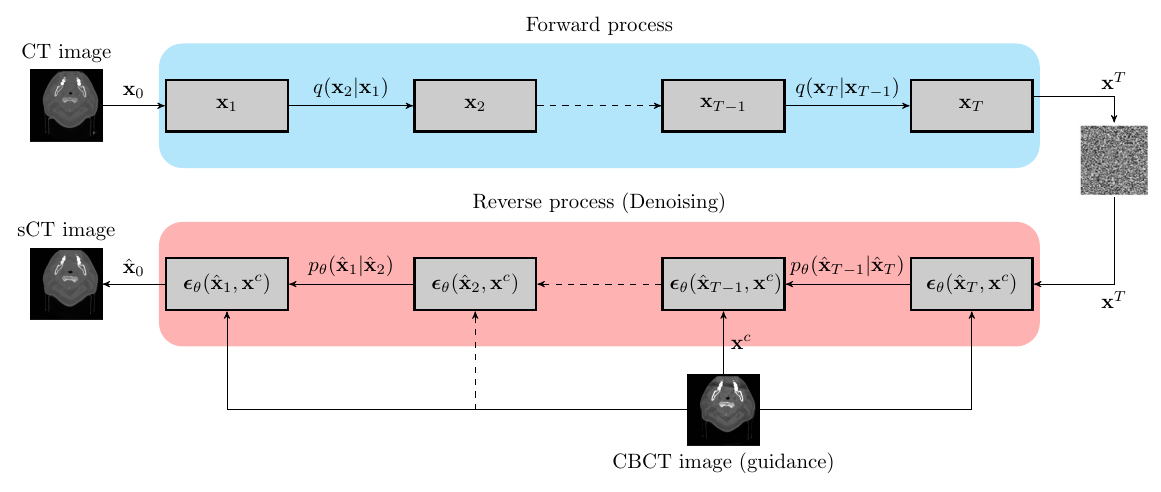}
	\caption{Workflow of EqDiff-CT for CBCT to CT image synthesis.}\label{fig:DDPM_workflow}
\end{figure}

The EqDiff-CT approach has been depicted in Fig.~\ref{fig:DDPM_workflow}  and allows them to reconstruct high quality sCT images with improved artefact reduction and HU accuracy.

The overall pipeline adopted in the study has been presented in Algorithm \ref{alg-cbct-ct}.

\begin{algorithm}[h]
\caption{CBCT to CT Synthesis Using Conditional Denoising Diffusion}
\label{alg-cbct-ct}
\begin{algorithmic}[1]
\REQUIRE Paired CBCT and CT images $\{\*x_i, \*x^c_i\}_{i=1}^N$, number of timesteps $T$, noise schedule $\{\beta_t\}_{t=1}^T$, conditional UNet model $\epsilon_{\bm\theta}$
\ENSURE Trained model $\epsilon_{\bm\theta}$ and synthesized CT $\hat{\*x}$

\vspace{0.5em}
\STATE \textbf{Initialize:} Compute $\alpha_t = 1 - \beta_t$, $\bar{\alpha}_t = \prod_{s=1}^t \alpha_s$
\FOR{each training epoch}
    \FOR{each minibatch $\{\*x^c, \*x\}$}
        \STATE Sample random timestep $t \sim \mathcal{U}(1, T)$
        \STATE Sample Gaussian noise $\boldsymbol{\epsilon} \sim \mathcal{N}(0, \mathbf{I})$
        \STATE Generate noisy CT: $\*x_t = \sqrt{\bar{\alpha}_t} \*x + \sqrt{1 - \bar{\alpha}_t} \boldsymbol{\epsilon}$
        \STATE Concatenate inputs: $\*z_t = [\*x_t, \*x^c]$
        \STATE Predict noise: $\hat{\boldsymbol{\epsilon}} = \epsilon_{\bm\theta}(\*z_t, t)$
        \STATE Compute loss: $\mathcal{L} = \text{MSE}(\hat{\boldsymbol{\epsilon}}, \boldsymbol{\epsilon}) + \lambda_{SSIM} \cdot (1 - \text{SSIM}(\hat{\bm\epsilon}, \bm\epsilon))$
        \STATE Update $\bm\theta$ using gradient descent
    \ENDFOR
\ENDFOR

\vspace{0.5em}
\STATE \textbf{Inference (Sampling from noise):}
\STATE Given CBCT input $\*x^c$ and initial noise $\*x_T \sim \mathcal{N}(0, \*I)$
\FOR{$t = T$ to $1$}
    \STATE Concatenate $\*x_t$ with CBCT: $\*z_t = \left[\*x_t, \*x^c\right]$
    \STATE Predict noise: $\hat{\boldsymbol{\epsilon}}_t = \epsilon_{\bm\theta}(\*z_t, t)$
    \STATE Estimate denoised image: 
    \[
    \*x_{t-1} = \frac{1}{\sqrt{\alpha_t}} \left( \*x_t - \frac{1 - \alpha_t}{\sqrt{1 - \bar{\alpha}_t}} \hat{\boldsymbol{\epsilon}}_t \right) + \sigma_t  \bm\epsilon, \quad \bm\epsilon \sim \mathcal{N}(0, \*I)
    \]
    \STATE Clip $\mathbf{x}_{t-1}$ to valid HU range $[-1000, 2000]$
\ENDFOR
\RETURN $\hat{\*x} = \*x_0$
\end{algorithmic}
\end{algorithm}

\subsection{Group Equivariant Conditional U-Net}\label{sec:equiv}
The intuition behind the idea of EqDiff-CT is that both the CT and CBCT physics acquisition rely on rotational measurements and this implies angular features or artifacts in the image domain. This is well known as most of the low dose CBCT images suffers from streaking artefacts with rotational direction respect to the centre of the scanner object. Therefore, EqDiff-CT builds upon the idea of designing rotational equivariant convolutional filters that can capture these angular features within the neural network and compensate for possible source of artifacts in the CBCT images respect to the reference CT pairs.

The group equivariant components used in EqDiff-CT are based on prior work on group equivariant convolutional neural networks and steerable equivariant CNNs. 
The denoising model $\epsilon_\theta$ in the proposed conditional DDPM framework has been developed using a group equivariant U-Net which follows the formulation of Cohen and Welling~\cite{cohen2016group}. The steerable E(2) equivariant convolutional \texttt{e2cnn} implementation follows the framework of Weiler and Cesa~\cite{weiler2019general}. In this work, we used the \texttt{e2cnn} PyTorch library~\cite{weiler2019general,e2cnnrepo} to implement C4 equivariant R2Conv layers, FieldTypes, GeometricTensors, equivariant normalization, and equivariant nonlinearities.  Unlike standard CNNs, which are only translation equivariant, our network achieves equivariance to discrete in plane rotations (cyclic group $C_4$). This ensures that feature maps and learned filters produce orientation consistent responses when the CBCT input is rotated by multiples of $90^\circ$. Each convolutional layer in the U-Net is replaced by an \texttt{R2Conv} from \texttt{e2cnn}. This constrains filters to transform according to representations of the cyclic group $C_4$. Our contribution is the integration of C4 equivariant convolutional blocks into a conditional DDPM based CBCT to CT synthesis framework. The aim is to investigate whether a rotation consistent denoising architecture can improve artifact suppression and anatomical preservation in head and neck CBCT to CT
synthesis. 

Let $G \in C_4$ act on $\mathbb{Z}^2$ by rotations. For a feature field $f:\mathbb{Z}^2 \to \mathbb{R}^{C_{\mathrm{in}}}$, define the joint action
\begin{equation}
[g \!\cdot\! f](\*z) \;=\; \rho_{\mathrm{in}}(g)\, f(R_g^{-1} \*z),\qquad \*z\in \mathbb{Z}^2
\end{equation}
where $R_g$ is the spatial action and 
$\rho_{\mathrm{in}}:G\to GL(\mathbb{R}^{C_{\mathrm{in}}})$ 
is a channel representation as direct sums of regular irreducible representations encoded by \texttt{e2cnn} FieldTypes.  

A linear map $\Phi$ is equivariant iff
\begin{equation}
\Phi[g \!\cdot\! f] \;=\; g \!\cdot\! \Phi[f] \qquad \forall g\in G .
\end{equation}

\noindent\textbf{Definition [Steerable ($\mathbb{R}^2$) Group Convolution]}: Let $\kappa:\mathbb{Z}^2 \to \mathrm{Hom}(\mathbb{R}^{C_{\mathrm{in}}},\mathbb{R}^{C_{\mathrm{out}}})$ 
be a steerable kernel satisfying the intertwining constraint
\begin{equation}
\kappa(R_g \*x) \;=\; \rho_{\mathrm{out}}(g)\,\kappa(\*z)\,\rho_{\mathrm{in}}(g)^{-1},
\qquad \forall g\in G,\, \*z\in \mathbb{Z}^2
\end{equation}

\noindent \textbf{Definition [Rotationally Equivariant \texttt{R2Conv} Block for $\mathbf{C_4}$]:} Given $f$, $\Phi$ and $\kappa$, the discrete equivariant convolution
\begin{equation}
[\Phi f](\*z) \;=\; \sum_{\*y\in\mathbb{Z}^2} \kappa(\*z - \*y)\, f(\*y)
\end{equation}
is $G$-equivariant and yields an output field with channel type $\rho_{\mathrm{out}}$. When $\rho_{\mathrm{out}}$ includes the regular representation, channels organize into $|G|$ orientation channels.

This constraint enforces rotation consistent responses across encoder and decoder stages. The integration of group equivariant convolutions reduces redundancy in learning rotated filters and improves generalization across patient orientation variability. 

\begin{figure}[!h]
\centering
\includegraphics[width=.8\linewidth]{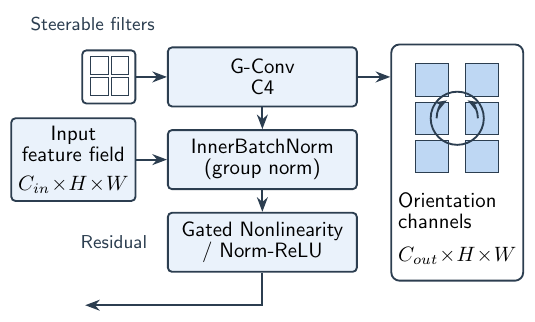}
\caption{Diagram of the equivariant \texttt{R2Conv} block used within the U-Net based network for the reverse diffusion process.}\label{fig:e2cnn}
\end{figure}

The equivariant output decompose into irreducible representations:
\begin{equation}
\rho_{\mathrm{out}} \simeq \bigoplus_i m_i\,\rho_i.
\end{equation}
and the \texttt{InnerBatchNorm} normalizes each irreducible block separately using a $G$-invariant inner product, preserving equivariance, while the \texttt{Norm-ReLU (gated)} acts on each block $\*v \in \mathbb{R}^{d_i}$ via
\begin{equation}
\mathrm{NormReLU}(\*v) \;=\; \frac{\sigma(\|\*v\|)}{\max(\|\*v\|,\varepsilon)}\, \*v,
\end{equation}
which is $G$-equivariant since it depends only on invariant norms. Fig.~\ref{fig:e2cnn} shows the overall \texttt{R2Conv} block constituted by the sub blocks described above. 

\usetikzlibrary{arrows.meta,positioning,calc,fit,backgrounds}
\definecolor{ink}{RGB}{44,62,80}      
\definecolor{tile}{RGB}{190,215,243}  
\definecolor{panel}{RGB}{234,242,251} 

\tikzset{
	>=Stealth,
	font=\sffamily,
	box/.style={draw=ink, rounded corners=2pt, line width=0.9pt, fill=panel, align=center},
	thinbox/.style={draw=ink, rounded corners=2pt, line width=0.9pt, align=center},
	arrow/.style={-Stealth, line width=0.9pt, draw=ink},
	labeltext/.style={font=\sffamily\small, text=ink},
}

The above definitions follow the standard group equivariant and steerable CNN
formulation. They are included here to clarify the notation used for the proposed
network, while the theoretical foundations and implementation are attributed to
the cited prior work~\cite{cohen2016group,weiler2019general,e2cnnrepo}. 
	
Combined with the conditional DDPM objective, this implementation stabilizes training and improves reconstruction fidelity in regions with high HU discontinuities such as the mandible and cervical spine.

\subsection{Time Conditioned Equivariant U-Net}\label{sec:time_equivariant_unet}
We adopt planar rotation equivariance with the cyclic group $G=C_4$. Input and output live in the trivial representation, while all hidden feature fields use regular representations. Convolutions are implemented with $G$-equivariant \texttt{R2Conv} layers, and normalization uses \texttt{InnerBatchNorm}; nonlinearities are \texttt{NormReLU}.

The time dependent denoiser in the reverse process is a three scale U-Net $\epsilon_{\bm\theta}$, which is parameterized by $\bm\theta$, with channel multipliers $\,(1,2,4)\,$. The encoder comprises three double conv blocks (each block: two $3{\times}3$ $G$-equivariant convolutions, each followed by 
\texttt{InnerBatchNorm} and \texttt{ReLU}), interleaved with $G$-equivariant down sampling. The decoder mirrors this structure using $G$-equivariant up sampling and skip connections to the corresponding encoder stages. We do not use an internal residual addition inside a block; the only residual connections are the U-Net skip connections across scales. This helps in preserving both local details and equivariant features. CBCT conditioning is concatenated channel wise at each resolution scale. This ensures that equivariant feature maps are modulated by anatomical contexts.

The model is explicitly conditioned on a diffusion time step $t \in \{0, \ldots, T-1\}$. This allows swift learning across noise levels. The model input comprises the concatenation of the noisy image $\*x_t \in \mathbb{R}^{1 \times H \times W}$ and the CBCT condition $[x_t \, \Vert \, x^{c}]$ which is then mapped by the \texttt{R2Conv} blocks. This leads to producing the predicted noise $\hat{\epsilon}_{\bm\theta}(\*x_t \| \*x^c, t)$, with a conditional time step index $t$. 

\paragraph{Time embedding and injection:} to inject temporal conditioning into the model, $t$ has been encoded into a continuous embedding vector $\boldsymbol{e}_t \in \mathbb{R}^{d}$. This is by using sinusoidal positional encoding, $p_t\!\in\!\mathbb{R}^{d}$ with components
\begin{equation}
	\*p^{\sin}_t[k] \;=\; \sin\!\left(\frac{t}{10000^{2k/d}}\right),\;
	\*p^{\cos}_t[k] \;=\; \cos\!\left(\frac{t}{10000^{2k/d}}\right)
\end{equation}
and set $\*p_t=[\*p^{\sin}_t; \*p^{\cos}_t]$. A two layer MLP maps $\*p_t$ to a learned embedding
\begin{equation}
	\*e_t = \phi \left( \mathbf{W}_
	{i+1} \cdot \sigma\left(\mathbf{W}_i \cdot \*p_t^\top \right) \right)
\end{equation}
where $\sigma(\cdot)$ denotes the ReLU activation function, and $\phi(\cdot)$ denotes the second ReLU transformation at block $i$. The projection weights $\mathbf{W}_i, \mathbf{W}_{i+1}$ are learned during training. Time is injected once at the bottleneck via a broadcast addition
\begin{equation}
	\*h_{i+1} \;\leftarrow\;
	\*h_i \;+\; \gamma(\*e_t),
\end{equation}
where $\gamma(\cdot)$ is a linear projection of the time embedding into the residual channel space.

\subsubsection{Residual Block with Temporal Conditioning and Channel Attention}

Each convolutional block in the encoder and decoder is implemented and represented as a residual unit:
\begin{align}
	\*h_i &= \texttt{GN}(\*z) \rightarrow \texttt{NormReLU} \rightarrow \texttt{R2Conv}(\mathbf{W}_i) \\
	\*h_{i+1} &= \mathbf{h}_i + \gamma(\boldsymbol{e}_t) \nonumber\\
	\*h_{i+1} &= \texttt{GN}(\mathbf{h}_{i+1}) \rightarrow \texttt{ReLU} \rightarrow \texttt{Dropout} \rightarrow \texttt{R2Conv}(\mathbf{W}_{i+1}) \nonumber\\
	\*y_{i+1} &= \texttt{Attn}(\mathbf{h}_{i+1} + \texttt{Shortcut}(\*z))  \nonumber
\end{align}
with $\texttt{GN}$ the group normalization \texttt{InnerBatchNorm}. The shortcut connection is a $1\times 1$ convolution if channel dimensions differ, or else are retained as identity.

\noindent The equivariant attention mechanism is designed using a self attention block with queries, keys, and values computed as:
\begin{align}
	\mathbf{Q} &= \mathbf{W}_q \mathbf{h}, \quad
	\mathbf{K} = \mathbf{W}_k \mathbf{h}, \quad
	\mathbf{V} = \mathbf{W}_v \mathbf{h} \\
	\*A_\star\,\bm\rho_{\mathrm{in}}(g) & =\bm\rho_{\mathrm{q/k/v}}(g)\,\*A_\star . \nonumber
\end{align}
where $\mathbf{W}_q, \mathbf{W}_k, \mathbf{W}_v$ are $1 \times 1$ convolutions. Using a $G$-invariant inner product $\langle\cdot,\cdot\rangle$ to define attention weights
\begin{equation}
\alpha_{rj} = \mathrm{softmax}_j\!\Big(\tfrac{1}{\sqrt{d}}\langle Q_r, K_j\rangle\Big)
\end{equation}
then 
\begin{equation}
\mathrm{Attn}(f)_r \;=\; \sum_j \alpha_{rj}\,V_j .
\end{equation}
is equivariant.

\textbf{Proposition:} If $\*W_q, \*W_k, \*W_v$ are intertwiners and $\langle\cdot,\cdot\rangle$ is $G$-invariant, then $\mathrm{Attn}$ is $G$-equivariant.

\subsubsection{Encoder and Decoder}

The U-Net has been structured in a hierarchical manner using the channel multipliers $[1, 2, 4]$ across four resolution scales. The encoder comprises a sequence of residual blocks that are followed by downsampling layers:
\begin{equation}
\*z_{i+1} = \texttt{DownSample}\Big(\texttt{ResBlock}(\*z_i, \*e_t)\Big)
\end{equation}
while the decoder replicated this pattern using upsampling and skip connections:
\begin{equation}
\mathbf{y}_{j-1} = \texttt{ResBlock}\Big(\texttt{Concat}[\mathbf{y}_j, \mathbf{x}_j], \boldsymbol{e}_t\Big) \rightarrow \texttt{UpSample}
\end{equation}

\subsubsection{Middle Bottleneck and Final Prediction}

At the lowest resolution, two central residual blocks have been developed. One with and one without attention, which serve as the bottleneck:
\begin{equation}
\*z_{i+1} = \texttt{ResBlock}_2\Big(\texttt{ResBlock}_1(\*z_i, \*e_t), \*e_t\Big)
\end{equation}
The output is then subjected to a final normalization, activation, and $3 \times 3$ convolution to generate the predicted noise:
\begin{equation}
\hat{\epsilon}_{\bm\theta}(\*x_t| \*x^c, t) = \texttt{R2Conv}\Big(\texttt{ReLU}(\texttt{GN}(\mathbf{y}_t))\Big)
\end{equation}

\begin{figure}[!h]
\centering
\includegraphics[width=1\linewidth]{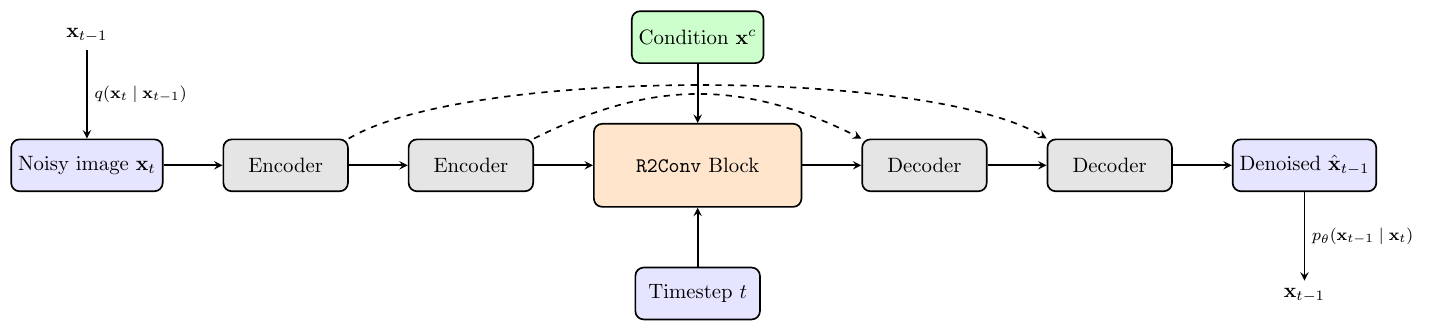}
	\caption{Diagram of the time dependent equivariant U-Net for step $t$ of the reverse process of EqDiff-CT.}\label{fig:EqDiff-CT}
\end{figure}

The overall diagram of the time Conditioned Equivariant U-Net at step $t$ in the reverse process of EqDiff-CT is shown in Fig.~\ref{fig:EqDiff-CT}.

\tikzstyle{arrow} = [->, thick, >=stealth]
\tikzstyle{residual} = [->, thick, dashed, >=stealth]
\tikzstyle{block} = [rectangle, draw, thick, minimum width=6em, minimum height=2.5em, align=center, rounded corners]

This architecture allows the model to learn the inverse diffusion using multiple noise levels and also preserves the anatomical structures and spatial consistency. Residual learning is integrated along with time conditioning and attention to generate powerful capacity. This leads to high fidelity medical image synthesis.

\subsection{Implementation: Dataset Design}\label{sec:dataset}

\subsubsection{Volumetric Data Acquisition and Organization}
The dataset employed in this study involves paired volumetric CBCT and planning CT images, collected for individual patients in three dimensional (3D) form. Let $\*x_i^c \in \mathbb{R}^{D \times H \times W}$ and $\*x_i \in \mathbb{R}^{D' \times H' \times W'}$ denote the volumetric CBCT and CT scans for the $i$-th patient, respectively. Due to inter patient variability in slice depth ($D$), spatial resolution, and field of view, all volumes undergo a standard preprocessing pipeline. This helps in ensuring uniformity.

The implementation design involves a complete pipeline, including data preprocessing, normalization, and sampling at the first stage for preparing  CBCT and CT slices. These are then subjected to the training of the diffusion based generative framework. The preprocessing protocol ensures that the inconsistencies associated with the geometry, intensity normalization, and anatomical alignments are addressed. These are essential for developing a stable convergence model for high capacity generative models in the clinical imaging context. 

\subsubsection{Cropping and Spatial Refinement}
To further eliminate the influence of non anatomical regions, each volume $\*x$ is cropped to the smallest subvolume $\*x_s$ that encloses all non zero voxels. Formally, we compute the bounding box $\mathcal{B} = \{\alpha, \beta, \gamma ~|~ \*x[\alpha, \beta, \gamma] > 0\}$ and extract:
\begin{equation}
\*x_s = \*x\left[\alpha_{\min}:\alpha_{\max},~ \beta_{\min}:\beta_{\max},~ \gamma_{\min}:\gamma_{\max}\right]
\end{equation}
where $\min$ and $\max$ are computed over the non zero support. This leads to improved anatomical centering and also improves the downstream contrast for normalization. 

\subsubsection{HU Normalization and Windowing}
Since both CBCT and CT data are represented in HU thus a fixed intensity window $[H_{\min}, H_{\max}] = [-1000, 2000]$  has been defined for standardization purposes across the two modalities. Each axial slice $S \in \mathbb{R}^{H \times W}$ is transformed using clipped min-max normalization into the canonical $[-1, 1]$ range:
\begin{equation}
\hat{S} = 2 \cdot \frac{\text{clip}(S, H_{\min}, H_{\max}) - H_{\min}}{H_{\max} - H_{\min}} - 1
\end{equation}
The transformation helps in preserving the tissue specific contrasts, including lung, bone, and soft tissue. This helps in mitigating the inter scan HU variability, which is a key limitation of CBCT. 

\subsubsection{2D Slice Extraction and Padding}

Each of the normalized volumes is segmented into axial slices. Since the heterogeneity in the slice dimensions exists due to cropping, the 2D slices $S_i$ are symmetrically padded with zeros to a target resolution $(H_t, W_t) = (224, 224)$:
\begin{equation}
S_i^{\text{pad}} = \text{Pad}\left(S_i, H_t, W_t\right)
\end{equation}
where $\text{Pad}(\cdot)$ performs zero padding in a way that spatial alignment is preserved. This leads to avoiding the ratio distortions. Such a fixed resolution further ensures that the compatibility is retained with the convolutional neural backbones. Subsequently, the diffusion pipelines are fed with uniform input size. 

\subsubsection{Data Augmentation and Pairing}
To improve the generalizability of the model, the training dataset incorporates stochastic augmentation schemes. These include horizontal and vertical flips as defined below:
\begin{equation}
S_i^{\text{aug}} = \begin{cases}
	\text{Flip}_x(S_i), & \text{if } r_1 > 0.5 \\
	\text{Flip}_y(S_i), & \text{if } r_2 > 0.5
\end{cases}
\quad \text{where } r_1, r_2 \sim \mathcal{U}(0,1)
\end{equation}
Each training sample consists of a paired tuple $\left(\hat{S}_{\text{CBCT}}^{(i)}, \hat{S}_{\text{CT}}^{(i)}\right)$ which represent the source and target domains. All slices are converted into single channel tensors $\mathbb{R}^{1 \times H_t \times W_t}$ deemed suitable for diffusion training.

\subsubsection{Dataset Splitting and Sampling Strategy}
The patient wise dataset is randomly shuffled and split into training and testing subsets in an 80/20 ratio. Let $\mathcal{D}_{\text{train}}$ and $\mathcal{D}_{\text{test}}$ denote the resulting sets:
\begin{equation}
\mathcal{D}_{\text{train}} = \{(\*x^c_i, \*x_i)\}_{i=1}^{N_{\text{train}}}, \quad
\mathcal{D}_{\text{test}} = \{(\*x^c_j, \*x_j)\}_{j=1}^{N_{\text{test}}}
\end{equation}
where $x_i$ and $y_i$ are CBCT and CT slices, respectively. The total number of 2D paired samples is computed as:
\begin{equation}
N = \sum_{p=1}^{P} \min\left(\text{depth}(\*x_p^c), \text{depth}(\*x_p)\right)
\end{equation}
These ensure that the slices are matched anatomically using the index across the modalities. 

\subsubsection{Image Intensity De normalization}

After generation, the synthetic outputs are mapped back to HU space by taking the inverse of the earlier normalization equation:
\begin{equation}
S_{\text{HU}} = \left(\frac{S_{\text{gen}} + 1}{2}\right) \cdot (H_{\max} - H_{\min}) + H_{\min}
\end{equation}
This allows for direct clinical interpretation, visualization, and integration with radiotherapy dose engines. 

\section{Results}\label{sec:results}
 
\subsection{Dataset Description and Training Parameters}
This study employed the publicly available synthRAD2025 dataset \cite{Thummerer2025SynthRAD2025}, focusing on the Head and Neck (HN) cohort, which is specifically curated to support research in CBCT to planning CT image synthesis for radiotherapy applications. The dataset comprises a total of 325 patient cases, yielding approximately 23,927 axial slices that encompass critical anatomical structures such as the mandible, airway, and cervical spine, areas of high clinical relevance in head and neck cancer treatment.

Each patient record includes paired CBCT volumes, acquired using low dose cone-beam imaging protocols, typically affected by increased noise, scattered artifacts, and reduced soft tissue contrast, and CT volumes, acquired using fan beam scan and used as ground truth for training and evaluation. All scans were processed into 2D axial slices with standardized dimensions of $224 \times 224$ pixels. 

The dataset was divided into training set containing 259 patients (approximately 80\%) and testing subsets of 66 patients (approximately 20\%) based on patient ID to prevent data leakage. For model training, both CBCT and CT image pairs were randomly sampled from the training set, while during testing, only CBCT images were provided as input, with the corresponding CT scans reserved for evaluation. This split ensures a robust and clinically realistic testing scenarios for synthetic CT generation tasks. To further clarify the inference protocol, no ground-truth CT image was provided to the model during testing or image generation. During inference, EqDiff-CT receives only the CBCT slice as the conditional input together with the stochastic reverse diffusion process. The corresponding CT image is used only after synthesis for retrospective quantitative evaluation and visual comparison. Therefore, the generated synthetic CT is conditioned on the CBCT image and does not directly access the reference CT during inference. The SynthRAD2025 HN clinical dataset is publicly accessible via GitHub at \cite{dataset} and the code of EqDiff-CT is available in the GitHub repository \url{https://github.com/ALZAHRAALTALIB/EqDiff-CT}.  

For EqDiff-CT algorithm we implemented rotational equivariance using the $C_4$ cyclic group ($0^{\circ}, 90^{\circ}, 180^{\circ}, 270^{\circ}$ rotations) in EqDiff-CT, rather than finer discretizations such as $C_8$ or continuous $SO(2)$. This choice was motivated primarily by computational considerations: $C_8$-equivariant training increased the per epoch runtime from approximately 26 minutes ($C_4$) to 1 hour 45 minutes ($C_8$), making extensive diffusion training impractical. Importantly, $C_4$ is sufficient to capture the dominant angular characteristics of CBCT artifacts arising from the acquisition geometry. CBCT streaking and scatter patterns exhibit approximately orthogonal orientations corresponding to the detector and gantry rotations; thus, enforcing equivariance at $90^{\circ}$ increments align with the underlying physics of artifact formation. 

The model has been trained used a batch size of $8$ with $650$ epochs. Th optimisation employed Adam solver with parameters $\gamma_1=0.9$, $\gamma_2=0.999$ and with a fixed learning rate of $10^{-4}$. The objective combined MSE and SSIM with equal weights $\lambda_{MSE} = \lambda_{SSIM} = 0.5$. Regarding the forward Diffusion model, $10^3$ step linear $\beta$-schedule was used with $\beta\in[10^{-4},  0.02]$. 

Different network configurations were investigated to assess the impact of the equivariant idea on the enhancements of sCT generation:
\begin{itemize}
\item \textbf{DDPM\cite{peng2024conditional} (Baseline):} this model replicates the conditional Denoising Diffusion models with U-Net architecture. 
\item \textbf{ACID-CT:} this configuration builds upon the baseline by integrating attention blocks at multiple stages of the U-Net, coupled with multi-GPU training. This design enhances both the convergence speed and the model's capacity to capture long range spatial dependencies. 
\item \textbf{Proposed EqDiff-CT:} This configuration builds upon ACID-CT by designing $C_4$ group convolutional blocks to enforce rotation consistency in the image domain. 
\item \textbf{CycleGAN \cite{liang2019}:} although it was originally proposed for unpaired image translation, we include it as a widely used baseline for CBCT to CT synthesis. We trained CycleGAN without pixel wise supervision, such that paired data were used only for evaluation, ensuring consistency with its original formulation.
\end{itemize}

The synthesized images $\hat{\*x}_0$ have been assessed compared to the original CT images $\*x_0$ using multiple image quality metrics using the Structural Similarity Index, $\text{(SSIM)}(\hat{\*x}_0, \*x_0) \in [0, 1]$ and the Peak Signal to Noise Ratio $\text{(PSNR)}(\hat{\*x}_0, \*x_0) = 10 \cdot \log_{10} \left( \frac{1}{\text{MSE}(\hat{\*x}_0, \*x_0)} \right)$ where $\text{(MSE)} = \frac{1}{N} \sum_{i=1}^{N} \left( \*x_0^{(i)} - \hat{\*x}_0^{(i)} \right)^2$. These metrics have been computed across all the slices in the subset to determine the spatial consistency and the inter slice findings.

\subsection{Ablation Study}

To investigate the contribution of the equivariance approach in the design of the UNet attention based network in the reverse process, we conduct an ablation study by comparing the EqDiff-CT model with the configuration without \texttt{e2cnn} block, i.e. the network constituted by attention blocks at multiple stages of the U-Net, coupled with multi-GPU training. In the following we will present a full set of quantitative and qualitative evaluation of the proposed EqDiff-CT method with alternative diffusion based network called ACID-CT (Attention based Conditional Image Diffusion CT) which does not use group equivariant convolutional $C_4$ blocks.

\subsubsection{Training Convergence}
\begin{figure}[!h]
    \centering
\includegraphics[width=.5\linewidth]{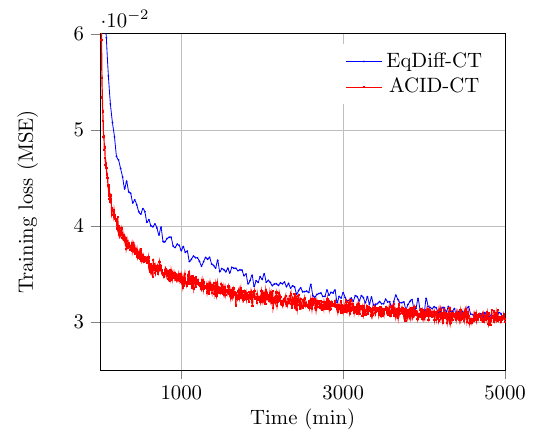}
    \caption{Training loss (MSE) over time (min) for EqDiff-CT and ACID-CT (model without equivariance).}\label{fig:loss}
\end{figure}

\begin{figure}[!h]
	\centering
	\includegraphics[width=\linewidth]{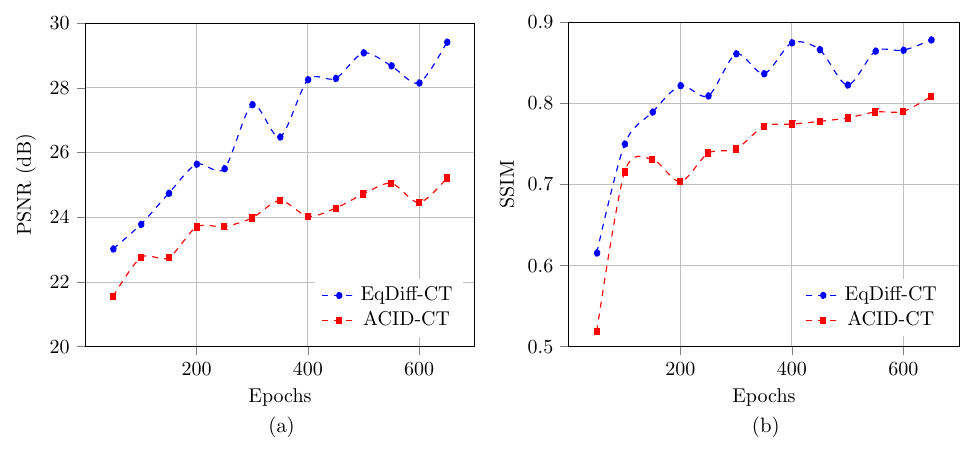}
	\caption{Ablation study for EqDiff-CT and ACID-CT (model without equivariance): (a) average PSNR (dB) over epochs and (b) average SSIM.} \label{fig:abl_psnr_ssim}
\end{figure}

Both EqDiff-CT and ACID-CT models were trained was over 650 epochs. Fig.~\ref{fig:loss} shows the training loss versus the actual time, while the EqDiff-CT module is slower at early iterations, it achieves the same average loss as the baseline network after the training period with less variability. Fig.~\ref{fig:abl_psnr_ssim} depict the PSNR and SSIM metrics versus the number of epochs, supporting the statement that the EqDiff-CT approach leads to consistent performance improvements compared to the baseline models without equivariance.

Using the checkpoint for the trained models obtained at 650 epochs, the results on the testing dataset (Table~\ref{tab:test-results}) show that the EqDiff-CT model consistently outperformed the baseline for the sCT image generation. In particular, the quantitative analysis shows that the average SSIM increased of 0.03 and reduced variance, indicating better structural preservation. Furthermore, the PSNR increased of around 0.9 dB and lower variance. 

These improvements validate the effectiveness of incorporating attention mechanisms in enhancing the perceptual quality and numerical stability of synthetic CT reconstruction. To note that since the validation dataset is consistent across simulations, the CBCT vs CT metrics are the same, indicating that observed improvements stemmed from model architecture, not input data variation. 

\begin{table}[h!]
	\centering
	\caption{Test Set Evaluation Metrics.}
	\label{tab:test-results}
	\begin{tabular}{ll|l>{\columncolor{Gray}}l}
		\toprule
		\hline
		\textbf{Comparison} & \textbf{Metric}         & \makecell{\textbf{ACID-CT}\\\textbf{w/o Equivariance}} & \makecell{\textbf{EqDiff-CT}\\\textbf{w/Equivariance}} \\
		\hline
		\midrule
		\multirow{2}{*}{CT vs Synth CT} 
		& SSIM           & 0.82 $\pm$ 0.10                 & $\mathbf{0.85} \pm \mathbf{0.09}$ \\         
		& PSNR (dB)      & 26.87 $\pm$ 4.25                  & $\mathbf{27.74} \pm \mathbf{3.98}$               \\
		\midrule
		\multirow{2}{*}{CBCT vs CT} 
		& SSIM           & \multicolumn{2}{c}{0.54 $\pm$ 0.14}  \\
		& PSNR (dB)      & \multicolumn{2}{c}{20.64 $\pm$ 4.35}                 \\
		\hline
		\bottomrule
	\end{tabular}
\end{table}

These findings highlight the ability of the equivariance approach in the EqDiff-CT model to capture the rotational correlations in the dataset and to quantitatively enhance the synthesis of CT images  form CBCT.

\subsection{Comparison EqDiff-CT (Equivariance) and ACID-CT (with Augmentation)}

A natural question is whether the performance gains of EqDiff-CT arise from the imposed rotational equivariance itself or could alternatively be achieved through conventional data augmentation. To investigate this, we compare the proposed equivariant diffusion model with a non equivariant counterpart trained using explicit rotation augmentation. EqDiff-CT consistently achieves superior performance as in Table \ref{tab:equiv_vs_aug} compared to the ACID-CT model with augmentation, $90^\circ$ rotations. 

\begin{table}[!h]
\centering
\caption{Comparison EqDiff-CT (Equivariance) and ACID-CT (with augmentation)}
\label{tab:equiv_vs_aug}
\begin{tabular}{l|>{\columncolor{Gray}}cc}
\hline
\rowcolor{green!20}
& \multicolumn{2}{c}{\textbf{CT vs Synth CT}} \\
\hline
\textbf{Metric} & \makecell{\textbf{EqDiff-CT}\\\textbf{(Equivariance)}} & \makecell{\textbf{ACID-CT}\\\textbf{(with augmentation)}} \\
\hline
\hline
SSIM            & $\mathbf{0.85} \pm \mathbf{0.09}$ & $0.70 \pm 0.14$  \\
PSNR (dB)       & $\mathbf{27.74} \pm \mathbf{3.98}$ & $22.44 \pm 4.71$  \\
\hline
\hline
\rowcolor{red!20}
& \multicolumn{2}{c}{\textbf{CBCT vs CT}} \\
\hline
SSIM & \multicolumn{2}{c}{$0.54 \pm 0.14$} \\
PSNR (dB) & \multicolumn{2}{c}{$20.64 \pm 4.35$} \\
\hline
\end{tabular}
\end{table}

Rotation augmentation encourages statistical invariance by exposing the network to transformed samples during training. However, the network must still learn the underlying symmetry from data, which can be inefficient and unreliable in limited or artifact corrupted datasets. In contrast, EqDiff-CT enforces exact transformation consistency at every layer by design, ensuring that intermediate feature representations transform predictably under rotations. This architectural constraint effectively encodes a structured inductive bias aligned with the angular characteristics of CBCT artifacts. 

We attribute this improvement to two factors. First, equivariance enforces transformation consistency at inference time, whereas augmentation only regularizes the training distribution. Second, equivariant convolutions reduce redundancy by sharing parameters across orientations, improving sample efficiency and stabilizing training in the presence of severe streaking and scatter artifacts. These advantages are particularly relevant for CBCT data, where artifact patterns exhibit structured angular dependencies that are difficult to learn implicitly.

\subsection{Comparison Computational Time EqDiff-CT and DDPM}

To compare to computational cost of EqDiff-CT and DDPM, we considered a representative case of one patient with 61 slice at testing. As shown in Table \ref{tab:compDDPM}, the DDPM required 38.4 min ($\approx 37.8$ s/slice, $1.6$ slices/min). EqDiff-CT, which contains rotational \texttt{R2conv} on top of UNet with attention modules, resulted in $34$ min ($\approx 33.5$ s/slice, $1.8$ slices/min), indicating no additional cost respect to DDPM. Thus, equivariance contributes negligible overhead in our implementationbase on NVIDIA GPU RTX A5000, FP32, batch size $=1$ and $250$ sampling steps at inference.)

\begin{table}[!h]
\centering
\caption{Inference runtime on a 61 slice case. Per slice latency and relative runtime are averaged over the single run shown; DDPM serves as the 1.0× reference.}\label{tab:compDDPM}

\begin{tabular}{lccc}
\toprule
Method & Total time (min) & Per slice (s) & Throughput  \\
& &  & (slices/min) \\

\midrule
DDPM & 38.4 & 37.8 & 1.6 \\
EqDiff-CT & 34.0 & 33.5 & 1.8 \\
\bottomrule
\end{tabular}
\end{table}

\subsection{Comparison with State of the art methods: DDPM and CycleGAN}

Following the ablation study, we assess the accuracy and robustness of EqDiff-CT by a comparative evaluation with a CycleGAN based approach \cite{liang2019}, a novel generative adversarial network (GAN) for image to image translation tasks and conditional Diffusion model DDPM. The CycleGAN model was configured with two standard U-Net based generator discriminator pairs and trained with a combination of adversarial, cycle consistency, and identity losses. Although CycleGAN is typically suited for unpaired image translation, it was adapted here to the paired setting for direct comparison. Both models were trained and tested on the same synthRAD2025 dataset to ensure a fair comparison.

\begin{figure}[!h]
	\centering
	\includegraphics[width=\linewidth]{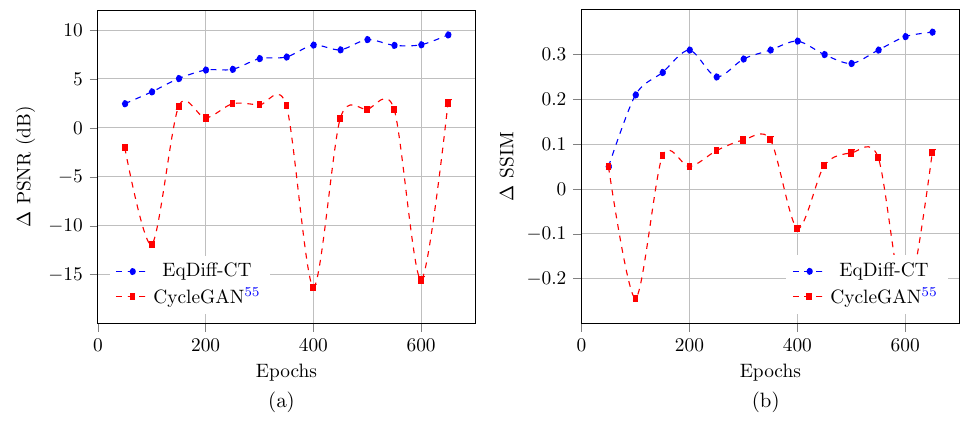}
	\caption{Improvement over epochs (CBCT comparison): (a) PSNR and (b) SSIM.} \label{fig:ssim_psnr_trend}
\end{figure}

First to analyse the consistency during training, we evaluated the SSIM and PSNR metric improvement over CBCT across epochs for both EqDiff-CT and CycleGAN. The equivariant diffusion model EqDiff-CT achieves increased improvement in the performance with smooth increase across epochs in all metrics as shown in Fig.~\ref{fig:ssim_psnr_trend}. For example, the PSNR achieved by EqDiff-CT at 650 epochs is around 8 dB higher compared to CycleGAN improvement and EqDiff-CT steadily increased across training, while CycleGAN showed high variability and instability.

We performed the same analysis on new clinical data at testing; Table~\ref{tab:cycle_vs_diffusion} summarizes the quantitative results in terms of average value and variance SSIM and PSNR metrics across the test dataset. EqDiff-CT model consistently outperformed CycleGAN and the Conditional Diffusion model DDPM across all metrics with decreased variability in the results. In particular EqDiff-CT improves the PSNR of around 3 dB in comparison with DDPM and 6.5 dB compared to CycleGAN.

\begin{table}[!h]
\centering
\caption{Average and Standard Deviation quantitative comparison: EqDiff-CT, DDPM, CycleGAN on Test Set.}
\label{tab:cycle_vs_diffusion}
\begin{tabular}{l|>{\columncolor{Gray}}ccc}
\hline
\rowcolor{green!20}
& \multicolumn{3}{c}{\textbf{CT vs Synth CT}} \\
\hline
\textbf{Metric} & \textbf{EqDiff-CT} & \textbf{DDPM} & \textbf{CycleGAN} \\
\hline
\hline
SSIM            & $\mathbf{0.85} \pm \mathbf{0.09}$ & $0.79 \pm 0.11$  &  $0.67 \pm 0.16$ \\
PSNR (dB)       & $\mathbf{27.74} \pm \mathbf{3.98}$ & $24.77 \pm 3.88$  &  $21.16 \pm 4.16$ \\
\hline
\hline
\rowcolor{red!20}
& \multicolumn{3}{c}{\textbf{CBCT vs CT}} \\
\hline
SSIM & \multicolumn{3}{c}{$0.54 \pm 0.14$} \\
PSNR (dB) & \multicolumn{3}{c}{$20.64 \pm 4.35$} \\
\hline
\end{tabular}
\end{table}

\begin{figure}[H]

	\begin{center}
\includegraphics[width=.95\linewidth]{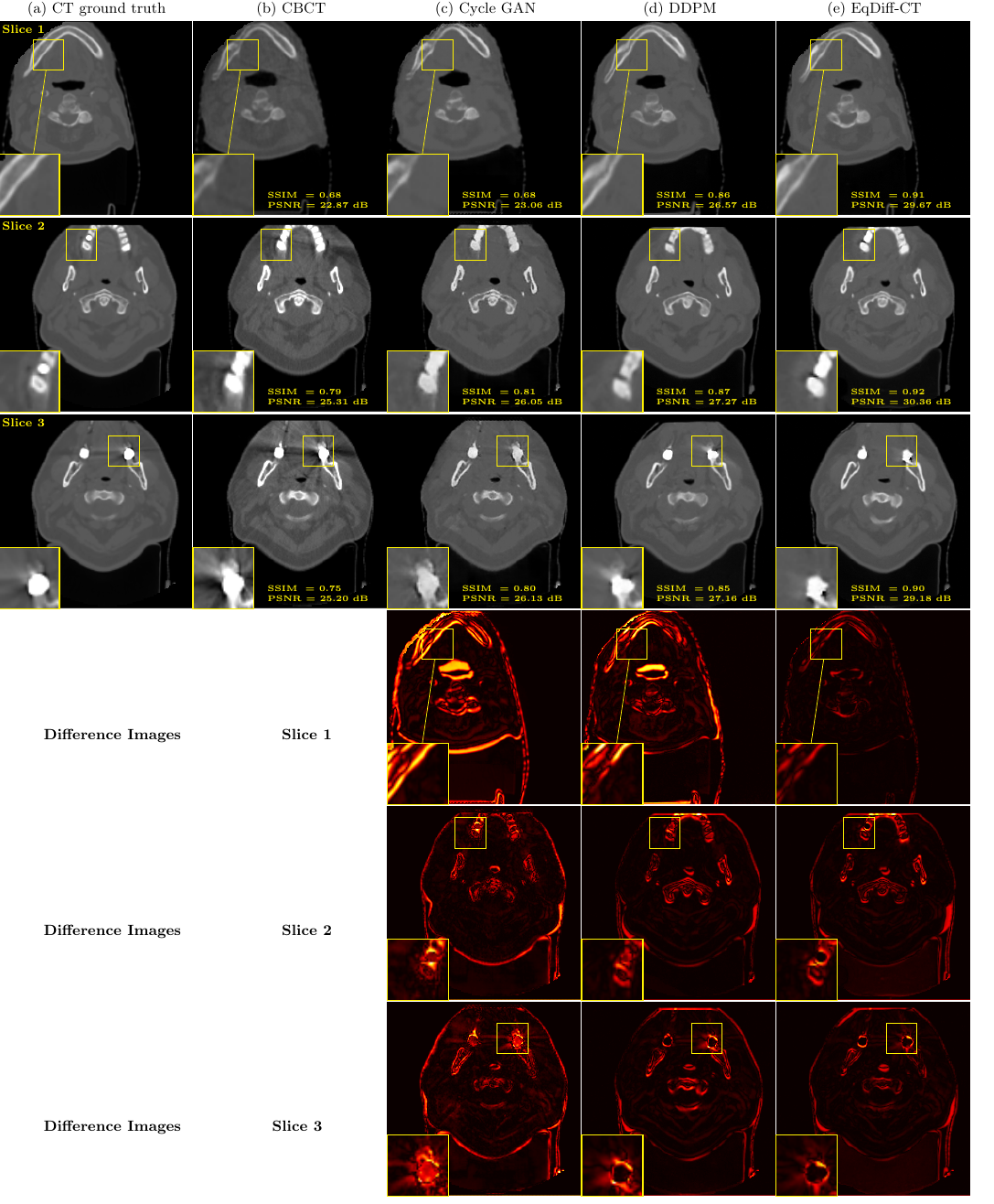}
    \end{center}
    \caption{Representative head and neck CBCT to synthetic CT results. comparison using contour consistent CBCT and CT slices. Columns show CT, CBCT, CycleGAN, DDPM, and EqDiff-CT, with zoomed regions highlighting local anatomical fidelity and artifact suppression. EqDiff-CT demonstrates improved anatomical consistency. Quantitative test set results show higher SSIM and PSNR for EqDiff-CT than competing methods. The reference CT is shown only for retrospective evaluation and was not used during testing or inference.}
\label{fig:results_compare}

\end{figure}

To evaluate the visual accuracy of the generated sCT images, qualitative comparisons were performed using representative axial slices selected to have visually consistent CBCT–CT external contours and anatomical alignment. This selection reduces ambiguity from registration or contour mismatch and allows the visual comparison to focus on artifact suppression, HU consistency, and anatomical preservation. As illustrated in Fig.~\ref{fig:results_compare}, the top row shows a) the original reference CT ground truth image, b) the corresponding CBCT input, c) sCT generated using CycleGAN, d) DDPM and e) EqDiff-CT, together with zoomed regions of interest. The bottom row represents the heat maps of the absolute difference between sCT and reference CT for both models. The reference CT is shown only for retrospective comparison and was not provided to the model during testing or inference. During inference, EqDiff-CT receives only the CBCT image as the conditional input; the CT image is used only after synthesis for visual comparison and quantitative evaluation. 

In Slice 1, the CBCT image contains local intensity non-uniformity near the highlighted high gradient boundary. CycleGAN preserves the general anatomy but provides limited improvement in structural similarity, while DDPM improves the CT like appearance and reduces part of the local intensity error. EqDiff-CT provides the highest quantitative agreement in this row, with SSIM increasing to 0.91 and PSNR increasing to 28.67 dB. The zoomed region shows that EqDiff-CT better preserves the local boundary structure and remains consistent with the CBCT supported contour, indicating that the improvement is mainly related to intensity correction and local artifact reduction rather than unsupported tissue generation.

In Slice 2, the highlighted region contains dense bony anatomy and strong bone / soft tissue contrast. The input CBCT shows reduced local contrast and artifact related HU inconsistency around the selected structure. CycleGAN improves the image appearance slightly but remains locally blurred, and DDPM provides better anatomical recovery but still shows residual intensity mismatch in the zoomed region. EqDiff-CT achieves the best performance in this row, with SSIM of 0.92 and PSNR of 30.36 dB. The zoomed comparison shows improved recovery of the high density structure and sharper local agreement with the reference CT, while maintaining the same visible anatomical contour as the CBCT input. 

In Slice 3, the highlighted region demonstrates the clearest example of structured streaking artifact reduction. The CBCT image contains visible rotational or streaking like intensity distortion around the high density region. CycleGAN and DDPM reduce some of the degradation but retain residual streak like patterns and local HU inconsistency. In contrast, EqDiff-CT substantially suppresses the residual artifact and provides the most CT local appearance, with SSIM of 0.90 and PSNR of 29.18 dB. This observation supports the role of the C4 rotationally equivariant module as an orientation consistent inductive bias for improving reconstruction stability in regions affected by structured CBCT artifacts.

Overall, EqDiff-CT produces anatomically sharper and more consistent reconstructions, particularly around clinically relevant high gradient structures such as the airway, mandible, cervical vertebral region, high density bony anatomy, and bone soft tissue or air tissue interfaces. The absolute HU error maps further demonstrate reduced reconstruction error with EqDiff-CT, especially near regions with abrupt attenuation changes. In particular, the mandible and cervical spine regions show lower residual intensity differences compared with CycleGAN and DDPM. 

The improvement observed in bony regions should be interpreted as better reconstruction of high gradient bony interfaces rather than direct correction of artifacts inside homogeneous bone tissue. In the zoomed regions of Fig. \ref{fig:results_compare}, EqDiff-CT shows sharper recovery of high density structures and lower residual HU error around bone soft tissue and bone air boundaries, including the mandible and cervical vertebral region. These interfaces are particularly sensitive to CBCT scatter, streaking  intensity variation, HU inconsistency, and partial volume effects. Compared with CycleGAN and DDPM, EqDiff-CT better preserves the local bony contour and reduces residual error around these artifact prone boundaries, supporting the role of C4 equivariance as an orientation consistent inductive bias for stabilizing reconstruction in high gradient anatomical regions. 

CycleGAN outputs were either overly smooth or exhibited local artifacts. The jawbone continuity and nasal cavity details were particularly better reconstructed by the diffusion model.

\begin{figure}[H]

	\begin{center}
\includegraphics[width=\linewidth]{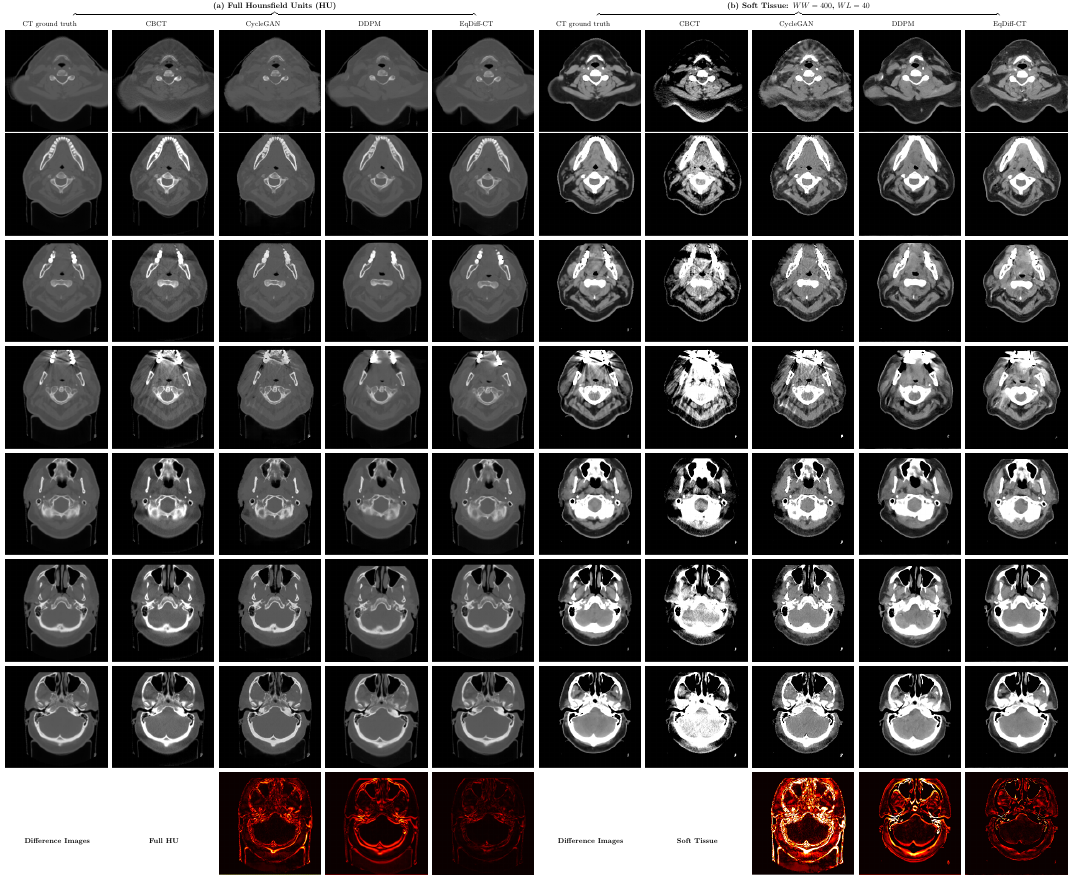}
     \caption{Representative qualitative comparison of CBCT to CT synthesis results for a head and neck case. Panel (a) shows images displayed using the full Hounsfield unit (HU) range, while panel (b) shows the same anatomical slices using a soft tissue window ($WW = 400$, $WL = 40$). For each panel, columns show the ground truth CT, input CBCT, CycleGAN result, DDPM result, and the proposed EqDiff-CT result. Multiple axial slices from different anatomical levels are included to evaluate structural consistency across the head and neck region. The bottom row shows absolute difference maps between the ground truth CT and the corresponding synthetic CT images, highlighting residual synthesis errors. Compared with the baseline methods, EqDiff-CT provides improved anatomical preservation and reduced error patterns, particularly around bone soft tissue interfaces and regions affected by CBCT artifacts.}\label{fig:res_soft_tissue}
\end{center}
\end{figure}

Figure~\ref{fig:res_soft_tissue} presents a qualitative comparison between the input CBCT, ground truth CT, and the synthetic CT images generated by CycleGAN, DDPM, and the proposed EqDiff-CT model for a representative head and neck case. The images are displayed using both the full HU range and a soft tissue window ($WW = 400$, $WL = 40$) to assess the preservation of global anatomical structure as well as subtle soft tissue contrast. Across different axial levels, the input CBCT shows visible intensity non uniformity and artifact related degradation, particularly around high density bone and air/tissue interfaces. 

The CycleGAN results reduce some CBCT artifacts but show local structural inconsistencies and residual intensity mismatch compared with the reference CT. 
The DDPM results provide improved anatomical appearance compared with CycleGAN; however, residual smoothing and local differences remain visible in several regions.  

In contrast, EqDiff-CT demonstrates closer visual agreement with the ground truth CT, with better preservation of anatomical boundaries and more consistent intensity appearance across slices. 
The absolute difference maps shown at the bottom of Figure~\ref{fig:res_soft_tissue} further highlight the spatial distribution of synthesis errors. 
Compared with the baseline methods, EqDiff-CT shows reduced error intensity in several artifact prone regions, especially near bone soft tissue boundaries and complex anatomical structures in the head and neck region. 
These qualitative findings support the quantitative results by showing that the proposed equivariant diffusion model improves CT like image synthesis from low quality CBCT while preserving clinically relevant anatomical details.

Furthermore, across all test dataset, while CycleGAN produced visually plausible results in some cases, its inconsistency, adversarial artifacts, and unstable metric trends limit its suitability for clinical workflows. EqDiff-CT, by contrast, demonstrated stable, high fidelity reconstruction with better generalizability across the test set. These findings reinforce the model’s potential for radiotherapy planning and other downstream clinical applications.

\subsection{Cross Domain Generalization of EqDiff-CT using Phantom Based Training}

To further evaluate the generalization capability of EqDiff-CT beyond a single training distribution, we conducted a cross domain experiment using a head dataset from a solid whole body phantom acquired using a varian CBCT scanner at Ninewells Hospital (Dundee, UK). Specifically, the model was primarily trained on CBCT–CT pairs acquired from a physical head phantom and subsequently fine tuned using a limited subset of the SynthRAD2025 dataset. The resulting model was then evaluated on the held out SynthRAD2025 test set. 

This experimental design aims to assess whether EqDiff-CT can leverage physics-driven artifact characteristics learned from phantom data and transfer them effectively to anatomically realistic patient images. Phantom CBCT data provide controlled acquisition conditions and well characterized artifact patterns, while the SynthRAD dataset introduces substantial anatomical variability and more realistic imaging noise.

\begin{table}[!h]
\centering
\caption{Comparison between Full training on SynthRad2025 and fine tuning using a phantom based training.}
\label{tab:fine_tuning}
\begin{tabular}{l|>{\columncolor{Gray}}cc}
\hline
\rowcolor{green!20}
& \multicolumn{2}{c}{\textbf{CT vs Synth CT}} \\
\hline
\textbf{Metric} & \makecell{\textbf{EqDiff-CT}\\\textbf{(Full Training)}} & \makecell{\textbf{EqDiff-CT}\\\textbf{(Fine tuning)}} \\
\hline
\hline
SSIM            & $\mathbf{0.85} \pm \mathbf{0.09}$ & $0.66 \pm 0.07$  \\
PSNR (dB)       & $\mathbf{27.74} \pm \mathbf{3.98}$ & $23.41 \pm 2.29$  \\
\hline
\hline
\rowcolor{red!20}
& \multicolumn{2}{c}{\textbf{CBCT vs CT}} \\
\hline
SSIM & \multicolumn{2}{c}{$0.54 \pm 0.14$} \\
PSNR (dB) & \multicolumn{2}{c}{$20.64 \pm 4.35$} \\
\hline
\end{tabular}
\end{table}

As shown in Table \ref{tab:fine_tuning}, EqDiff-CT trained under this protocol achieves competitive performance on the SynthRAD test set (3847 images), despite limited exposure to patient specific anatomy during training. This indicates that the model does not simply memorize anatomical appearance but instead learns transferable representations of CBCT artifact structure.  

\begin{figure}[H]

	\begin{center}
\includegraphics[width=\linewidth]{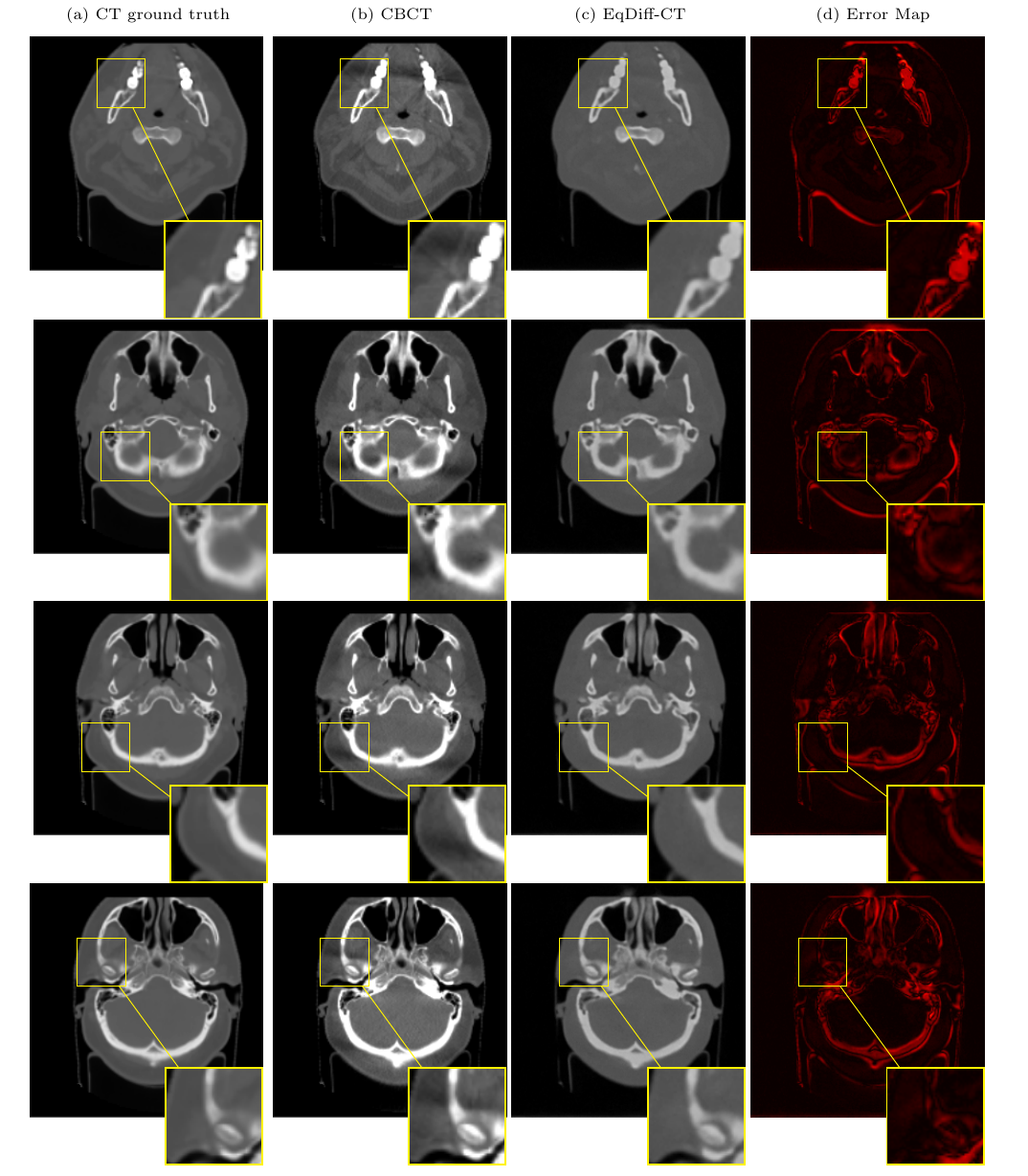}
     \caption{Qualitative CT image synthesis testing results and error map using EqDiff-CT cross domain training with phantom based dataset and SynthRAD2025 fine tuning: (a) CT, (b) CBCT, (c) EqDiff-CT and (d) error map between CT and EqDiff-CT.}\label{fig:result_pretrain}
\end{center}
\end{figure}

The ability to generalize from phantom based training to patient data supports the effectiveness of the proposed equivariant diffusion framework and its robustness to domain shifts in anatomy and acquisition conditions as shown in the sCT image and error map in Fig. \ref{fig:result_pretrain} obtained at testing with EqDiff-CT fine tuning training.

Although the fine tuned model improves the CT like appearance compared with the original CBCT input, the generated images appear smoother than those obtained from full training on SynthRAD2025. This residual smoothing is likely due to the limited patient specific anatomical variability available during the phantom based training stage and the remaining domain gap between phantom and clinical CBCT data. Therefore, this result demonstrates the transferability of the proposed framework under constrained training conditions. 

Overall, these results demonstrate that EqDiff-CT maintains strong synthesis performance under heterogeneous data settings, highlighting its potential for robust CBCT to CT synthesis in scenarios where large, fully paired clinical datasets are unavailable.

\section{Discussion}\label{sec:discussion}

\subsection{Main Findings}

This study introduced EqDiff-CT, a conditional diffusion framework for synthetic CT generation from CBCT images in head and neck radiotherapy. The main finding is that adding rotational equivariance to an attention based enhanced diffusion model improves the quality of CBCT to CT synthesis.

Compared with the non equivariant ACID-CT model, EqDiff-CT achieved higher SSIM and PSNR, indicating better structural preservation and improved intensity agreement with the reference CT.

These results suggest that equivariant feature learning can improve the ability of diffusion models to correct CBCT artifacts and recover CT like anatomical details.    

\subsection{Role of Rotational Equivariance} 

The improvement obtained by EqDiff-CT can be explained by the angular nature of CBCT artifacts. CBCT images are affected by scatter, streaking artifacts, and HU inconsistency, many of which are related to rotational acquisition geometry. By using C4 group equivariant convolutions, the model is encouraged to learn features that respond consistently to in plane rotations. This provides a useful architectural constraint, allowing the network to represent orientation dependent artifacts more effectively than a standard convolutional model. 

The revised qualitative comparison in Fig. \ref{fig:results_compare} further supports this interpretation. In Slice 3, streaking rotational artifacts remain visible in the CBCT image and are only partially reduced by CycleGAN and DDPM. EqDiff-CT substantially suppresses these residual streaking patterns and shows improved local HU consistency in the corresponding zoomed region. This observation supports the contribution of the C4 equivariant module as an orientation consistent architectural bias for improving reconstruction stability in regions affected by structured CBCT artifacts. However, C4 equivariance should not be interpreted as a complete physics based correction of scatter or streak artifacts.

The comparison with rotation augmentation also supports this interpretation. Although augmentation exposes the model to rotated images during training, it does not guarantee rotation consistent internal features. In contrast, equivariant convolutions enforce this consistency directly within the network. Therefore, the observed improvement is likely due to the equivariant design itself rather than only to increased data variation.

\subsection{Comparison with Existing Models} 

EqDiff-CT also outperformed the baseline DDPM and CycleGAN models. This is important because CycleGAN based methods can generate visually plausible images but may suffer from over smoothing or local hallucination artifacts. The diffusion based approach provides a more stable generation process by gradually denoising the image while remaining conditioned on the CBCT input. The addition of attention and rotational equivariance further improved anatomical preservation, especially around clinically important regions such as the mandible, airway, and cervical spine. 
Importantly, EqDiff-CT does not increase the inference cost compared with the standard DDPM baseline in a meaningful way. The equivariance constraint is imposed through the network design and optimized during training; no additional equivariance loss, registration step, or physics based constraint is computed during inference. As a result, EqDiff-CT uses the same reverse diffusion sampling procedure as DDPM, while benefiting from the rotation consistent features learned during model optimization. In the representative 65 patients test dataset with total pairs slices 3844, EqDiff-CT achieved a runtime comparable to DDPM, confirming that the proposed equivariant design does not introduce a prohibitive computational burden. Although diffusion based synthesis remains slower than single pass CNN or GAN methods, future work can reduce the overall sampling time using accelerated samplers, fewer denoising steps, latent diffusion, pruning, or knowledge distillation.

\subsection{Comparison with SynthRAD2025 CBCT to CT Studies (Task 2)} 

To better position EqDiff-CT with respect to existing SynthRAD2025 Task 2 literature, we compared our results with recent methods evaluated for CBCT to CT generation. Table \ref{tab:comp_synthrad2025} summarizes the reported Task 2 results from the selected studies from the SynthRAD2025 challenge. Where available, only the head and neck results are reported. However, some studies reported Task 2 CBCT to CT validation results across the validation set without separating anatomical regions; therefore, these comparisons should be interpreted cautiously.

\begin{table}[!h]
\centering
\caption{Comparison with selected SynthRAD2025 Task 2 CBCT to CT studies.  When available, head and neck (HN) results are shown.}\label{tab:comp_synthrad2025}
\label{tab:synthrad2025_task2_comparison}
\resizebox{\textwidth}{!}{
\begin{tabular}{llcc}
\toprule
Method & Reported setting & Region & PSNR $\uparrow$ \\
\midrule
3D U-Net \cite{ronneberger2015, mei2025} & Task 2 - CBCT to CT & All regions 
& $25.842 \pm 2.104$ \\
SwinUNETR \cite{hatamizadeh2021, mei2025} & Task 2 - CBCT to CT & All regions 
& $26.315 \pm 2.148$ \\
3D U-Net++ \cite{zhou2018, mei2025} & Task 2 - CBCT to CT & All regions 
& $26.754 \pm 2.205$ \\
Flow Matching \cite{hadzic2025} & Task 2 - CBCT to CT & All regions 
& $26.30 \pm 1.61$ \\
DynUNet \cite{Almasni2025} & Task 2 - CBCT to CT & HN 
& $26.48$ \\
Standard DDPM & Internal SynthRAD2025 test split & HN 
& $24.77 \pm 3.88$ \\
\textbf{EqDiff-CT} & \textbf{Internal SynthRAD2025 test split} & \textbf{HN} 
& $\mathbf{27.74 \pm 3.98}$ \\
\bottomrule
\end{tabular}
}
\end{table}

The comparison shows that several recent SynthRAD2025 Task 2 methods rely on 3D volumetric CNN architectures evaluated \cite{mei2025}, including 3D U-Net \cite{ronneberger2015}, SwinUNETR \cite{hatamizadeh2021}, 3D U-Net++ \cite{zhou2018} and DynUNet \cite{Almasni2025}. These methods benefit from volumetric context and, in some cases, challenge specific optimization. In contrast, EqDiff-CT focuses on a different contribution: improving diffusion based CBCT to CT synthesis through attention and C4 rotational equivariance. Under our internal SynthRAD2025 HN split, EqDiff-CT improved PSNR from $25.84 \pm 2.1$ dB with 3D U-Net to $27.74 \pm 3.98$ dB. It is worth noting that EqDiff-CT outperforms similar diffusion conditional models like Flow Matching CBCT to CT \cite{hadzic2025}. 

Although direct numerical comparison should be interpreted with caution because of differences in evaluation protocol, dimensionality, and reported metrics, the table clarifies the position of EqDiff-CT relative to recent SynthRAD2025 Task 2 studies. The proposed method demonstrates that equivariant diffusion modelling can improve over standard DDPM and U-Net networks for head and neck CBCT to sCT synthesis.

\subsection{Clinical Relevance}

The proposed method has potential value for CBCT guided adaptive radiotherapy. CBCT is routinely acquired during treatment, but its use for dose calculation and adaptive planning is limited by poor HU accuracy and image artifacts. By generating CT like images from CBCT, EqDiff-CT may support improved anatomical assessment, auto segmentation, and future dose recalculation workflows. The qualitative results showed improved reconstruction of bone and soft tissue boundaries, which is particularly important in head and neck radiotherapy where small anatomical errors may affect organs at risk and treatment evaluation.  

\subsection{Limitations and Future Work} 

This study has several limitations. First, the current implementation is based on 2D axial slices rather than full 3D volumes, which may limit inter slice consistency. Future work should investigate 3D \cite{altalib2026epc} or 2.5D versions of EqDiff-CT. Second, the evaluation was focused on the head and neck cohort, and further validation is required for other anatomical sites such as pelvis, thorax, and abdomen. Third, the method depends on the quality of CT and CBCT pairing and preprocessing alignment. Imperfect pairing, residual registration mismatch, acquisition differences, may lead to external contour inconsistencies. Fourth, future studies should focus on downstream tasks such as dose calculation, dose volume histogram analysis, gamma evaluation, and expert clinical review.

In addition, the current equivariant design used the C4 rotation group for computational efficiency. Future work may explore higher order equivariance, projection domain physics constraints and multi institutional validation. These steps are necessary before the proposed method can be considered for routine clinical implementation.

\section{Conclusion}\label{sec:conclusion}
This work proposes an equivariant diffusion model, called EqDiff-CT, for the synthesis of high fidelity CT images from CBCT. The aim was to address the challenges associated with the artifact correction and structural preservation. This framework is based on the intuition that noise artefacts belong to geometrical location related to the rotational direction of the CT acquisition setup. This idea is combined with generative learning by leveraging the inherent stability DDPM and employed a U-Net based architecture at the baseline for making accurate transformations across noise and low quality CBCTs. The model has been trained and evaluated on the SynthRAD2025 dataset. This involved the head and neck anatomy and depicted notable improvement in the image similarity metrics. The quantitative and qualitative results show that EqDiff-CT improves CT like image similarity, reduces residual intensity errors, and better preserves clinically relevant anatomical structures compared with the evaluated baseline methods. In addition, EqDiff-CT improves in recovering the anatomical details across the clinically relevant subregions. These include the brainstem, parotid glands, and mandible. The model can preserve the spatial fidelity and rectify the intensity based distortions. This makes it promising to downstream the tasks that include dose recalculation and auto segmentation. 

Overall, the work highlights a strong potential for the equivariant diffusion models to serve as a robust and generalizable approach in clinical settings for sCT generation compared to the state of the art alternative models such as CycleGAN and baseline diffusion model DDPM. In the future, we will be focusing on the integration of uncertainty quantification, multi organ generalization, and clinical validation on real world raw CBCT and CT datasets.

\subsection*{Author Contribution Statement}
Alzahra Altalib performed the Conceptualization, Methodology, Data curation, Formal analysis, Investigation and Writing, original draft and review. Chunhui Li contributed to the Supervision and Writing, review \& editing. Alessandro Perelli contributed to the Conceptualization, Methodology, Supervision and Writing,  review \& editing. All authors reviewed, edited, and approved the final version of the manuscript.

\subsection*{Acknowledgement Statement}
The authors thank Sandra and Yamen Alhyari for their contribution to improve the manuscript and Christopher Taylor and Sankar Pillai for supporting in the phantom data collection at Ninewells hospital. 

\subsection*{Funding}
A. Altalib is supported by the Jordan University of Science and Technology PhD scholarship. A. Perelli is supported by the Royal Academy of Engineering / Leverhulme Trust Research Fellowship LTRF-2324-20-160.

\subsection*{Conflict of Interest Statement}
The authors declare no conflicts of interest, whether financial or personal, that could influence or be relevant to the work presented in this paper.

\subsection*{Ethical statement} 
This work involve human subjects in its research. Approval of all ethical and experimental procedures was granted by the respective institutional review boards and medical committees of the SynthRAD2025 challenge.

\section*{References}
\addcontentsline{toc}{section}{\numberline{}References}
\vspace*{-10mm}

\bibliographystyle{medphy.bst}
\bibliography{refs}

\end{document}